\documentclass{aa}
\usepackage{times}
\usepackage{graphicx}
\usepackage{array}
\usepackage{epsfig}
\usepackage{amssymb}
\usepackage{amsmath}
\usepackage{longtable}
\usepackage{subfigure}

\begin{document}

\newcommand{\Mdot}{$\dot{M}$}
\newcommand{\Msun}{$\rm{M}_{\odot}$}
\newcommand{\Minit}{$M_{\rm init}$}
\newcommand{\Mwd}{$M_{\rm{WD}}$}
\newcommand{\Rsun}{$\rm{R}_{\odot}$}
\newcommand{\msyr}{$\rm{M}_{\odot}\rm{yr}^{-1}$}
\newcommand{\rtrip}{$3\alpha$}
\newcommand{\as}{$\alpha_{\rm s}$}
\newcommand{\apmath} {$\left(\frac{\partial \ln \rho}{\partial \ln P}\right)_T$}
\newcommand{\atmath} {$-\left(\frac{\partial \ln \rho}{\partial \ln T}\right)_P$}
\newcommand{\ap} {$\alpha_P$}
\newcommand{\cp} {$c_P$}
\newcommand{\at} {$\alpha_T$}
\newcommand{\DR} {$D/r_{\rm s}$}
\newcommand{\dad} {$\nabla_{\rm ad}$}
\newcommand{\sige} {$\sigma_{\rm E}$}
\newcommand{\D}{\displaystyle}
\newcommand{\density}{${\rm g~cm^{-3}}$}

\authorrunning{Yoon, Langer and van der Sluys}

\title{On the stability of thermonuclear shell sources in stars}

\author{S.-C. Yoon, N. Langer \and M. van der Sluys}

\institute{
Astronomical Institute, Utrecht University, Princetonplein 5,
NL-3584 CC, Utrecht, The Netherlands
}

\offprints {S.-C. Yoon  (email: {\tt S.C.Yoon@astro.uu.nl})}
\date{Received/Accepted}

\abstract{
We present a quantitative criterion for the thermal stability of thermonuclear 
shell sources. 
We find the thermal stability of shell sources to depend on exactly three
factors: they are more stable when they are geometrically thicker, 
less degenerate and hotter. 
This confirms and unifies previously obtained
results in terms of the geometry, temperature and density of the shell source,  
by a simplified but quantitative approach to the physics of shell nuclear burning.
We present instability diagrams in the temperature-density plane for
hydrogen and helium shell burning, which allow a simple evaluation of
the stability conditions of such shell sources in stellar models. 
The performance of our stability criterion is demonstrated in various
numerical models: in a 3 \Msun{} AGB star, in
helium accreting CO white dwarfs,  in a helium white dwarf which is 
covered by a thin hydrogen envelope,  and in a 1.0~\Msun{} giant.
}

\maketitle

\section{Introduction}
Thermonuclear fusion is the main source of energy in stars. To use fusion as
main energy supply over long time scales requires 
extreme stability, since the thermonuclear reaction rates are sensitive
functions of the temperature. I.e.,
a slight temperature increase may enhance the nuclear burning rate drastically;
if this results in an even higher temperature, a thermal runaway occurs.
In most stars, such a runaway does not occur, since an increased energy output
due to an enhancement of the nuclear burning rate leads to an overpressure and
thus to an expansion of the burning region (Kippenhahn \& Weigert~\cite{Kippenhahn90}). 
However, two different circumstances have been shown to be 
able to prevent such a pressure increase.
 
The first one is electron degeneracy. If nuclear burning
starts in a degenerate gas, the increased energy production will lead to a higher
temperature. But as --- for complete degeneracy --- the pressure does not depend
on the temperature, it remains constant. This situation is realized during the
core helium flash in low mass stars, at carbon ignition in Chandrasekhar-mass
CO white dwarf, i.e. in a Type~Ia supernova explosion, and in strong nova shell
flashes. 

The second circumstance is a strong geometrical confinement of the
burning region. As discovered by Schwarzshild \& H\"arm (\cite{Schwarzshild65}) 
and Weigert (\cite{Weigert66}),
geometrically thin helium shell sources in red giants can become unstable. 
Since the local gravity in a thin shell source remains constant even when it
expands in response to an increased energy production, hydrostatic equilibrium
will enforce a constant pressure in the shell. The thin shell instability 
is prone to occur in nuclear shell sources around compact cores, as the small
scale heights induced by the compactness of the core ensures the geometrical
confinement of the shells. Examples are thermal pulses in asymptotic giant branch
stars (e.g. Iben \& Renzini~\cite{Iben83a};  Busso et al.~\cite{Busso01}; Lugaro et al.~\cite{Lugaro03}), 
hydrogen or helium shells on white dwarfs 
(e.g. Fujimoto~\cite{Fujimoto82a},~\cite{Fujimoto82b}; Cassisi et al.~\cite{Cassisi98};
Langer et al.~\cite{Langer02}; Yoon et al.~\cite{Yoon04c}),
and X-ray bursts on neutron stars (e.g. Fujimoto et al.~\cite{Fujimoto81}; 
Taam \& Woosley~\cite{Taam96}).  

In principle, both destabilizing circumstances may occur together. 
In central burning regions, however, the degeneracy aspect dominates, as 
gravity and thus pressure will not be constant in an expanding core. 
In nuclear shell sources, degeneracy and geometrical confinement 
may simultaneously contribute to the 
instability of shell sources. Among various authors investigating the
stability of nuclear shell sources 
(e.g. Giannone \& Weigert~\cite{Giannone67}; Unno~\cite{Unno70}; 
Dennis~\cite{Dennis71}; H\"arm \& Schwarzshild~\cite{Harm72};  
Stothers \& Chin~\cite{Stothers72}; Sackmann~\cite{Sackmann77}), 
Kippenhahn \& Weigert~(\cite{Kippenhahn90})
developed a semi-quantitative understanding of the relevance of both destabilizing
factors. 

The purpose of this paper is to develop the insight into the
shell stability mechanism further such that
a quantitative tool is obtained, which 
allows to assess the thermal stability of shell sources in numerical stellar models.
This will allow for a physical interpretation of fluctuating nuclear shell sources
found in non-linear time-dependent stellar evolution calculations. 
As described in Section~2, we essentially follow the method by 
Giannone \& Weigert (\cite{Giannone67}),  but assume homology in the region of the shell 
source as in Kippenhahn \& Weigert (\cite{Kippenhahn90}).
In Section~3, we derive thermodynamic parameters for partial degenerate
conditions, which we use in Section~4 to formulate and investigate the 
stability criterion in idealized situations. In Section~5,
we apply the criterion to various numerical stellar models, and give a summary
and discussion of our results in Section~6.

\section{Secular behavior of shell burning}\label{sect_sec}

\subsection{Gravothermal specific heat}\label{sect_pertb}

Let us consider a shell source 
with a geometrical thickness $D$. The mass of the shell source is given as 
$\Delta M_{\rm s}=\int_{r_0}^{r_{\rm s}} 4\pi r^2 \rho dr$, where
$r_0$ is the radius of the bottom of the shell source and $r_{\rm s}=r_0 + D $, 
its upper 
boundary. Assuming $r_0$ is constant, 
 the relation between density and radius perturbation
is given from $d\Delta M_{\rm s}=0$ as 
\begin{equation}\label{eq0}
\frac{\delta\rho}{\rho}
=-\frac{3}{3D/r_{\rm s}-3(D/r_{\rm s})^2+(D/r_{\rm s})^3} \frac{\delta r_{\rm s}}{r_{\rm s}}
\end{equation}
(Huang \& Yu~\cite{Huang98}).
If we assume that the matter in the shell source expands or contracts
homologously, we have $\delta P/P = -4 \delta r/r$ from the hydrostatic equation 
and we obtain:
\begin{equation}\label{eq1}
\frac{\delta P}{P}=\alpha_{\rm s}\frac{\delta \rho}{\rho}
\end{equation}
where
\begin{equation}\label{eq2}
\alpha_{\rm s} = \frac{4}{3}\left(3D/r_{\rm s}-3(D/r_{\rm s})^2+(D/r_{\rm s})^3\right).
\end{equation}
Kippenhahn \& Weigert's value  \as=4\DR{} is restored when \DR$\ll 1$, while
it becomes 4/3 with \DR=1, 
which is the case when the whole star contracts or expands homologously.

Using the thermodynamic relation $\delta\rho/\rho = \alpha_P\delta P/P - \alpha_T\delta T/T$,
it follows from Eq.~(\ref{eq1}) that
\begin{equation}\label{eq3}
\frac{\delta P}{P} = \frac{\alpha_{\rm s}\alpha_{T}}{\alpha_{\rm s}\alpha_{P}-1}\frac{\delta T}{T}, 
~~~~~
\frac{\delta \rho}{\rho} = \frac{\alpha_{T}}{\alpha_{\rm s}\alpha_{P}-1}\frac{\delta T}{T}
\end{equation}
where $\alpha_P=(\partial\ln\rho/\partial\ln P)_T$ and  $\alpha_T=-(\partial\ln\rho/\partial\ln T)_P$
 (Kippenhahn \& Weigert~\cite{Kippenhahn90}).
As we will see in the next section, \at{} and \ap{} are functions of the degree of degeneracy
and are not constant over a shell source. 
However,
since we are interested in the mean properties of a shell source in this study,
we will assume that like the density and the pressure perturbation,
the temperature change is also homologous in the shell,
and mean values over the shell source 
will be used for \ap{} and \at.

The heat perturbation in the shell source can be obtained from 
the first law of the thermodynamics as shown by Kippenhahn \& Weigert (\cite{Kippenhahn90}):
\begin{equation}\label{eq4}
\delta q=c^*\delta T
\end{equation}
with
\begin{equation}\label{eq5} 
c^*:=c_{\rm P}\left(1 - \nabla_{\rm ad}\frac{\alpha_{\rm s}\alpha_{T}}{\alpha_{\rm s}\alpha_{P}-1}\right) ,
\end{equation}  
where \cp{}  denotes the specific heat at constant 
pressure and $\nabla_{\rm ad}=(d\ln T/d\ln P)_{\rm ad}$.

The quantity $c^*$ is called the gravothermal specific heat (Kippenhahn \& Weigert~\cite{Kippenhahn90}).
When the heat $\delta q$ is added to the shell source, 
more energy than $\delta q$ is consumed
by the expansion work if $c^* < 0$, thereby decreasing the internal energy. 
However, if $c^*>0$, the additional energy heats up the matter and thus favors
instability.  

\subsection{The stability criterion}\label{sect_crit}

The gravothermal specific heat alone does not describe 
the thermal reaction of a shell source, as it is
also influenced by its thermal interaction with its surroundings.
The energy conservation in a star is expressed by the following equation:
\begin{equation}\label{eq6}
\frac{\partial L_r}{\partial M_r} = \epsilon_{\rm N} 
- \frac{dq}{dt}.
\end{equation}
%Here we did not consider neutrino emission.
For a shell source with the thickness $D=r_{\rm s}-r_0$ and
the mass $\Delta M_{\rm s}$,
integration over the burning shell gives
\begin{equation}\label{eq7}
L_{r_{\rm s}} - L_{r_0} = L_{\rm N} - L_{\rm g}.
\end{equation}
$L_{r_{\rm s}}$ and $L_{r_0}$ are the luminosities at $r=r_{\rm s}$ and $r=r_0$, respectively. 
$L_{\rm N}$ is the luminosity due to the nuclear burning in the shell: 
$L_{\rm N}=\int_{\Delta M_{\rm s}} \epsilon_{\rm N} dM_r$ where $\epsilon_{\rm N}$ is the
nuclear energy generation rate in the shell source.
$L_{\rm g}$ is defined as
$\int_{\Delta M_{\rm s}} \frac{dq}{dt} dM_r$. 
In AGB stars or accreting white dwarfs 
the main contribution to $L_{\rm g}$ is due to the gravitational
energy release by contraction. 
This term is not significant compared to $L_{\rm N}$ in general,
and we will  assume that the shell source is initially in 
a stationary state (i.e., $dq/dt=0$).

Perturbing Eq.~(\ref{eq7}) yields
\begin{equation}\label{eq8}
 \delta L_{r_s}
= \Delta M_{\rm s}\delta\epsilon_{\rm N} - \Delta M_{\rm s}\frac{d\delta q}{dt}
\end{equation}
where $L_{r_0} \ll L_{r_s}$ has been assumed.
With the use of equations~(\ref{eq1})---(\ref{eq5}) and 
with assuming radiative heat transfer, i.e.,
\begin{displaymath}
\frac{\partial T}{\partial M_r} = -\frac{3}{64\pi^2ac}\frac{\kappa L_r}{r^4 T^3}, 
\end{displaymath} 
we obtain the following equation for the temperature perturbation 
$\theta=\delta T/T$:
\begin{equation}\label{eq9}
\tau_{\rm th}\dot{\theta}
=   \sigma\theta,
\end{equation}
with
\begin{equation}\label{eq10}
\tau_{\rm th}=\frac{\Delta M_{\rm s}Tc_{\rm P}}{L_{r_s}},
\end{equation}
and
\begin{equation}\label{eq11}
\sigma 
 = \frac{\D \nu - 4 + \kappa_{\rm T} + \frac{\alpha_T}{\alpha_{\rm s}\alpha_P - 1}
(\lambda + \alpha_{\rm s} + \kappa_{\rho})}
{\D c^*/c_p} .
\end{equation}
Here,  
$\kappa_{\rm T}=(\partial\ln\kappa/\partial\ln T)_\rho$,
$\kappa_{\rm P}=(\partial\ln\kappa/\partial\ln \rho)_T$, 
$\nu = (\partial\ln\epsilon/\partial\ln T)_\rho$, 
and $\lambda = (\partial\ln\epsilon/\partial\ln \rho)_T$.
The quantity $\tau_{\rm th}$ defined in Eq.~(\ref{eq10}) corresponds to 
the thermal time scale of the shell source. Note that Eq.~(\ref{eq9})
is essentially the same as the equation~(16)  in Giannone \& Weigert (\cite{Giannone67}), 
for the case that the assumption
of homology in the shell source is adopted in their analysis.

When $\sigma>0$ the shell source becomes thermally unstable and 
the perturbation growth time scale becomes
\begin{equation}\label{eq12}
\tau_{\rm growth} = \tau_{\rm th}/\sigma.
\end{equation} 
The larger the value of $\sigma$, the more rapidly 
the thermal instability grows. 
If $\sigma < 0$, nuclear burning in a shell source is thermally stable, and,
upon a temperature increase, the temperature drops faster for larger $\sigma$.
Therefore, $\sigma$ 
can serve as a measure of the susceptibility of a shell source to thermal instability.

For the following discussions, let us define the numerator in Eq.~(\ref{eq11}) as \sige{} such that
$\sigma=\sigma_{\rm E}/(c^*/c_P)$ and
\begin{equation}\label{eq13}
\sigma_{\rm E} = \nu - 4 + \kappa_{\rm T} + \frac{\alpha_T}{\alpha_{\rm s}\alpha_P - 1}
(\lambda + \alpha_{\rm s} + \kappa_{\rho}).
\end{equation}
The first 3 terms in the above equation are related to the temperature perturbation 
and the remaining terms to the density perturbation.

For a given $c^*$, i.e. a given equation of state, the stability of a shell source
is determined by the sign of \sige{}.
Physically,  $\sigma_{\rm E} > 0$ means 
that the additional energy production
in response to a positive temperature perturbation  
exceeds the additional energy loss by radiation. 
In other words, for $\delta T >0$, we have
$ \delta q > 0 $ if $\sigma_{\rm E} > 0$, in which case
a shell source is stable if $c^*<0$,  since, 
the additional heat 
is consumed mostly for the expansion work of the shell source 
as discussed in section~\ref{sect_pertb}. 
If $\sigma_{\rm E} < 0$, 
the additional energy loss by radiation  is larger than the additional energy production 
(i.e., $\delta q < 0)$ in response to $\delta T >0$. 
If $c^*<0$ and $\sigma_{\rm E} < 0$, this results in further increase in the temperature perturbation
since  the net energy loss is compensated by the contraction of the system,
increasing the internal energy of the shell source. 
However, this instability mode has little relevance as in stars 
\sige{} hardly becomes negative if $c^*<0$, as we will see in Sect.~\ref{sect_c_neg}. 
If $c^*>0$ and $\sigma_{\rm E} < 0$,  the net energy loss results in the
decrease in the internal energy and stable burning is established. 

In the following sections, we will investigate the possible modes
of stability/instability of a shell source 
in a more quantitative way.

\section{Physical conditions for stability/instability}\label{sect_physcond}

In order to discuss the stability of shell source in realistic stars,
it is necessary to understand the behaviour of the thermodynamic quantities
such as \ap, \at{} and $c^*$ in various equation-of-state regimes. 

\begin{figure*}[t]
\center
%\epsfxsize=0.8\hsize
%\epsffile{fig_thermo.eps}
\resizebox{0.8\hsize}{!}{\includegraphics{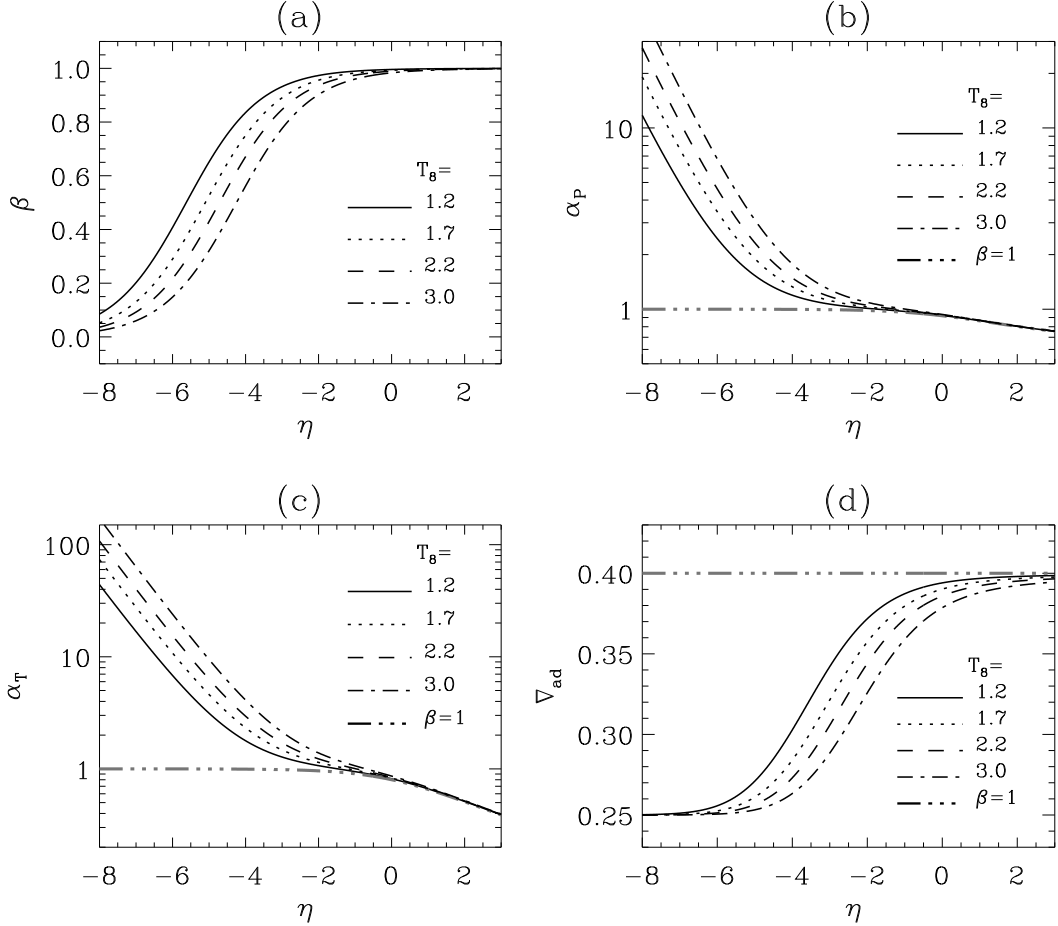}}
\caption{Thermodynamic quantities as a function of the degeneracy parameter $\eta$ 
for 4 different temperatures as indicated. Here, $T_8$ denotes temperature in unit of $10^8\rm{K}$. 
As a reference, 
the case of $\beta=1$ is also shown  for \ap, \at{} and $\nabla_{\rm ad}$.
(a) - Ratio of gas pressure to total pressure, $\beta$, as a function of the degeneracy parameter.
(b), (c) - $\alpha_P (:=(\partial \ln\rho/\partial\ln P)_T)$  
and $\alpha_T (:=(\partial \ln\rho/\partial \ln T)_P)$ calculated from Eq.~(\ref{eq19}) and~(\ref{eq20}) using
$\beta$ given in (a). 
(d) - $\nabla_{\rm ad} (:=(\partial \ln T / \partial \ln P)_{\rm ad})$  
calculated from Eq.~(\ref{eq22}) using $\beta$ and $\alpha_T$ given
in (a) and (c) respectively.  
}\label{fig_thermo}
\end{figure*}

For non-relativistic partially degenerate conditions, 
electron density and pressure are given by
\begin{equation}\label{eq14} 
n_{\rm e}  =  \frac{\rho}{\mu_{\rm e}m_{\rm H}}  =  \frac{4\pi}{h^3}(2m_{\rm e}kT)^{3/2}F_{1/2}(\eta) 
\end{equation}
and 
\begin{equation}\label{eq15}
P_{\rm e}  =  \frac{\rho kT}{\mu_{\rm e}m_{\rm H}}\frac{2F_{3/2}(\eta)}{3F_{1/2}(\eta)} 
\end{equation}
where  $F_{3/2}$ and $F_{1/2}$ are the Fermi-Dirac functions 
and $\eta:=\psi/kT$  is the degeneracy parameter (e.g. Clayton~\cite{Clayton68}). 
The equation of state  is thus given as:
\begin{equation}\label{eq16}
P = P_{\rm g}+P_{\rm r}= \frac{\rho kT}{\mu_{\rm I}m_{\rm H}}
+\frac{\rho kT}{\mu_{\rm e}m_{\rm H}}\frac{2F_{3/2}(\eta)}{3F_{1/2}(\eta)}
+\frac{1}{3}aT^4.
\end{equation}
With the definition of $\beta:=P_{\rm g}/P$, the above equation can be rewritten as
\begin{equation}\label{eq17}
F(\eta) + \frac{\mu_{\rm e}}{\mu_{\rm I}} = \frac{\mu_{\rm e}m_{\rm H}\beta P}{\rho kT} ,
\end{equation}
where $F:=2F_{3/2}/3F_{1/2}$. 
Differentiating this equation, we get
\begin{equation}\label{eq18}
\frac{F}{F+\mu_{\rm e}/\mu_{\rm I}}d\ln F = \frac{1}{\beta}d\ln P - \frac{4-3\beta}{\beta}d\ln T - d\ln\rho
\end{equation}
where $(\partial\ln\beta/\partial\ln P)_T=(1-\beta)/\beta$ and $(\partial\ln\beta/\partial\ln T)_P=4(\beta-1)/\beta$
were used. 
From Eq.~(\ref{eq18}) and by differentiating Eq.~(\ref{eq14}), we can obtain \ap{} and \at{} as a function of $\eta$:
\begin{equation}\label{eq19}
%\normalsize
\alpha_P = \frac{\D \frac{1}{\beta}\frac{d\ln F_{1/2}}{d\eta}}
{\D \frac{d\ln F_{1/2}}{d\eta} + \frac{F}{F+\mu_{\rm e}/\mu_{\rm I}}\frac{d\ln F}{d\eta} }
\end{equation} 
and
\begin{equation}\label{eq20}
\alpha_T = \frac{\D \frac{4-3\beta}{\beta}\frac{d\ln F_{1/2}}{d\eta} - \frac{3}{2}
\frac{F}{F+\mu_{\rm e}/\mu_{\rm I}}\frac{d\ln F}{d\eta}}
{\D \frac{d\ln F_{1/2}}{d\eta} + \frac{F}{F+\mu_{\rm e}/\mu_{\rm I}}\frac{d\ln F}{d\eta}}.
\end{equation} 
When $\beta=1$, we get \ap=\at=1 for an ideal gas 
($\eta\rightarrow-\infty$, Fig.~\ref{fig_thermo}), 
and \ap=0.6 and \at=0 
for complete degeneracy ($\eta\rightarrow\infty$). 
In the case of $\beta \not= 0$, 
we have $\alpha_P = 1/\beta$ and $\alpha_T = (4-3\beta)/\beta$ in a non-degenerate gas 
(cf. Kippenhahn \& Weigert 1990),
since $F(\eta) \rightarrow 1$ and $d\ln F/d\eta \rightarrow 0$ when  $\eta \rightarrow -\infty$.

The specific heat at constant pressure is given 
by $c_{\rm P} = \left(\frac{\partial u}{\partial T} \right)_P
-\frac{P}{\rho^2}\left(\frac{\partial\rho}{\partial T}\right)_P$ (Kippenhahn \& Weigert~\cite{Kippenhahn90}). 
From $u=\left(\frac{3}{2} P_{\rm g} + 3P_{\rm r}\right)/\rho = (3-\frac{3}{2}\beta)P/\rho$, we have
\begin{equation}\label{eq21}
c_{\rm P} = \frac{P}{\rho T} \left[ \left(\frac{5}{2} + \frac{3}{2}(1-\beta)\right)\alpha_T + 6(1-\beta) \right]
\end{equation}
and 
\begin{equation}\label{eq22}
\nabla_{\rm ad} = \frac{P\alpha_T}{\rho T c_{\rm P}} 
= \frac{\alpha_T}{\left( \frac{5}{2} + \frac{3}{2}(1-\beta)\right)\alpha_T + 6(1-\beta) }.
\end{equation}
Since $P/\rho T$ in Eq.~(\ref{eq21}) can also be given as a function of $\eta$ from Eq.~(\ref{eq17}),
the gravothermal specific heat $c^*$ 
can be obtained as a function of $\eta$ for a specified $\beta$.

The value of $\beta$ 
can be obtained as a function of $\eta$ at a given temperature from the equation of state.  
In Fig.~\ref{fig_thermo}a, we  show $\beta$ as a function of $\eta$ for 4 different temperatures 
which may be relevant to the helium shell burning. $\mu_{\rm e}=2$ and
$\mu_{\rm e}/\mu_{\rm I}=0.38$ have been assumed. In this temperature range, 
$\beta$ reaches 1 when $\eta \gtrsim 0$ since the gas
pressure becomes dominant.

As discussed in Sections~\ref{sect_c_neg}--\ref{sect_ills},
the values of \at{} and \ap{} and the sign of $c^*$ 
affect the stability of a shell source critically. 
Fig.~\ref{fig_thermo}b and~\ref{fig_thermo}c show 
\ap{} and \at{} as a function of $\eta$, obtained 
using the values of $\beta$ given in Fig.~\ref{fig_thermo}a. 
The case when the radiation pressure
is neglected (i.e., $\beta=1$) is also shown in these figures.
Since changes in the total pressure 
become more sensitive to changes in temperature than in density
as radiation pressure becomes significant, we find larger values
of \ap{} and \at{} when radiation pressure is considered than in the case of $\beta=1$.
Since \dad{} varies little compared to \ap{} and \at{},
the sign of $c^*$ is mainly determined by the sign of $\alpha_{\rm s}\alpha_P -1$ (see Eq.~(\ref{eq5})).

\begin{figure}[t]
%\epsfxsize=\hsize
%\epsffile{fig_grth.eps}
\resizebox{\hsize}{!}{\includegraphics{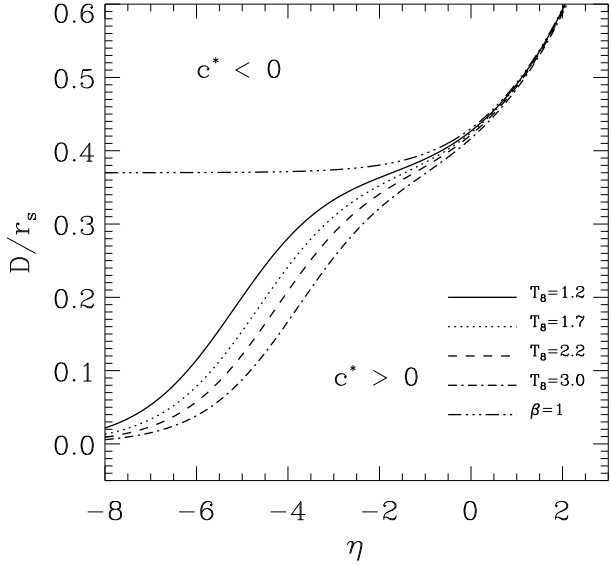}}
\caption{Lines of vanishing gravothermal specific heat $c^*$
in the plane spanned by the degeneracy parameter $\eta$ and the relative
thickness of the shell source $D/r_{\rm s}$, for four different
temperatures. These lines are computed by using the thermodynamic quantities
given in Fig.~\ref{fig_thermo}.   
The line labeled $\beta=1$ shows the case where radiation pressure
is neglected. 
}\label{fig_grth}
\end{figure}

Figure~\ref{fig_grth} shows the lines of zero gravothermal specific heat $c^*$,
in the plane spanned by the degeneracy parameter $\eta$ and
the relative thickness of the shell source $D/r_{\rm s}$, for 4 different temperatures. 
These lines have been determined using the values of $\beta$, \at, \ap{} and \dad{}
given in Fig.~\ref{fig_thermo}. 
For $\beta=1$, the thickness of the shell source 
should be larger than $0.37r_{\rm s}$ for the gravothermal specific heat to be negative 
in the non-degenerate region ($\eta \lesssim -2$). 
As the electron degeneracy becomes more significant, the minimum thickness which
gives $c^* < 0$ becomes larger. 

In the non-degenerate region, $c^*$ remains negative for much smaller
shell thicknesses when the effect of the radiation pressure
is considered. This is due to the fact that \ap{} and \at{}  
become large when radiation pressure is significant.
Since shell sources are
generally stable when $c^*<0$ as discussed in Sect.~\ref{sect_c_neg},
this indicates that radiation pressure serves as a stabilizing factor
for a shell source.
The stabilizing effect of radiation pressure, which has already been discussed
by many authors (e.g. Dennis~\cite{Dennis71}; Stothers \& Chin~\cite{Stothers72}; Sackmann~\cite{Sackmann77}),
is due to the fact that a small increase in temperature
induces considerable expansion of the shell source
when radiation pressure is significant (cf. Eq.~(\ref{eq3})).

\subsection{The case of $c^*<0$}\label{sect_c_neg}

When $c^*<0$, the following two conditions are fulfilled from Eq.~(\ref{eq5}):
\begin{displaymath}
\alpha_{\rm s}\alpha_P -1 > 0 ~~~ 
{\rm and} ~~~ \alpha_{\rm s}\alpha_P-1 < \nabla_{\rm ad}\alpha_{\rm s}\alpha_T ~~.
\end{displaymath}
The latter condition can be rewritten as $\alpha_P - \nabla_{\rm ad}\alpha_T < 1/\alpha_{\rm s}$.
In a non-degenerate gas (i.e., $d\ln F/d\eta \rightarrow 0$), we have
$\alpha_P - \nabla_{\rm ad}\alpha_T \rightarrow 0.75$ from Eqs.~(\ref{eq19}), 
(\ref{eq20}) and (\ref{eq22}), while
we have $\alpha_P - \nabla_{\rm ad}\alpha_T \rightarrow 0.6$ in a completely degenerate gas
where $\alpha_P = 0.6$ and $\alpha_T = 0$. It follows that the inequality  
$\alpha_{\rm s}\alpha_P-1 < \nabla_{\rm ad}\alpha_{\rm s}\alpha_T$ is always 
satisfied for all realistic values of $\alpha_{\rm s}$. 
Therefore, the condition  $c^*<0$ reduces to $\alpha_{\rm s}\alpha_P -1 > 0 $. 
With this condition, the term $\alpha_T/(\alpha_{\rm s}\alpha_P-1)$ in Eq. (\ref{eq13})
is positive,
and it is generally $\sigma_{\rm E} > 0$ 
since $\nu$ dominates over $\kappa_T$ and $\kappa_\rho$ in most cases.

\begin{figure}[t]
%\epsfxsize=\hsize
%\epsffile{fig_nu_3alpha.eps}
\resizebox{\hsize}{!}{\includegraphics{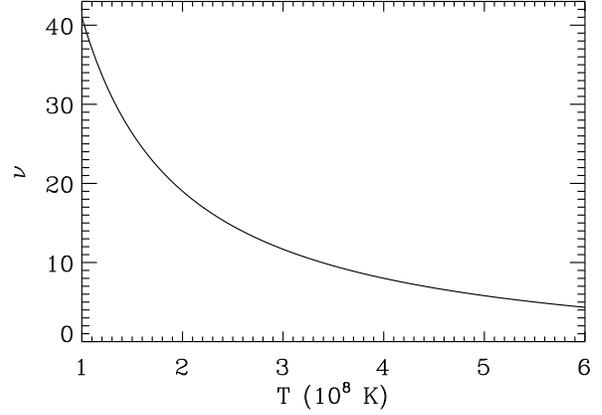}}
\caption{ 
The logarithmic derivative of the $3\alpha$ reaction with respect to temperature:
$\nu=(\partial\ln\epsilon_{3\alpha}/\partial\ln T)_{\rho}$.
The reaction rate given by Harris et al.~(\cite{Harris83}) has been used in the 
calculation. 
}\label{fig_nu_3alpha}
\end{figure}

\begin{figure}[t]
%\epsfxsize=\hsize
%\epsffile{fig_kapa_deriv.eps}
\resizebox{\hsize}{!}{\includegraphics{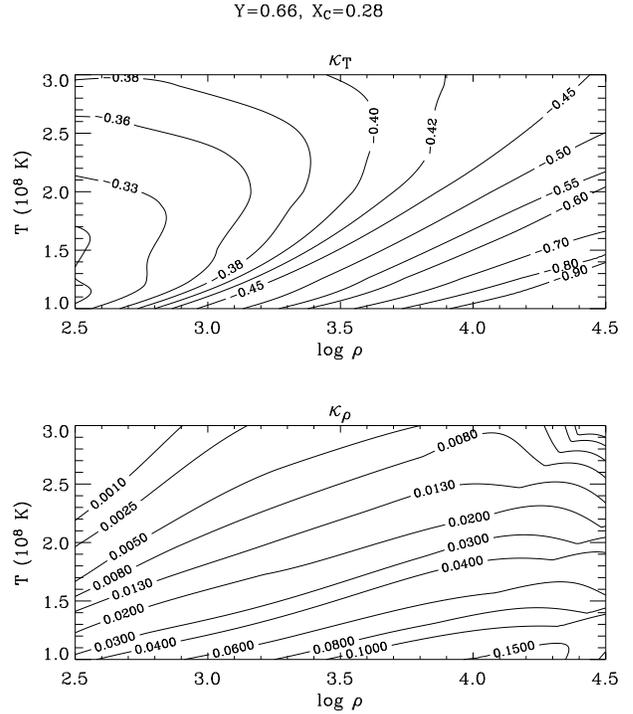}}
\caption{ The logarithmic derivatives of the opacity with respect to
to temperature and density: $\kappa_T=(\partial\ln\kappa/\partial\ln T)_{\rho}$
and $\kappa_\rho=(\partial\ln\kappa/\partial\ln \rho)_T$. 
The OPAL opacity table by Iglesias \& Rogers~(\cite{Iglesias96}) has been used in the calculation. 
$Y=0.662$, $X_{\rm C}=0.286$ and $Z = 0.02$  have been assumed. 
}\label{fig_kapa_deriv}
\end{figure}

For example, let us consider a helium shell source powered by
the $3\alpha$-reaction.
Figure~\ref{fig_nu_3alpha} shows $\nu$ 
as a function of temperature, calculated from the $3\alpha$ reaction rate given 
by Harris et al.~(\cite{Harris83}).
The density dependence $\lambda$ is equal to 2 regardless of temperature. 
In AGB stars or in accreting white dwarfs, the temperature of the helium shell source
varies according to the core mass or the accretion rate within the range
$1 < T/10^8~K < 3$, giving $\nu=40 \cdots 12$.
Figure~\ref{fig_kapa_deriv} shows the contour levels of $\kappa_T$ and $\kappa_{\rho}$
in the $T-\rho$ plane, calculated from the OPAL opacity table by Iglesias \& Rogers~(\cite{Iglesias96}).  
The helium abundance has been assumed to be 0.66 in this calculation since 
the energy generation reaches its maximum around this value in a helium shell source.
This figure shows that $\kappa_T>-1$
and $\kappa_{\rho}\ll 1$ 
in the given temperature and density range. 
Therefore, $\sigma_{\rm E} > 0$ is guaranteed as long as $\alpha_T/(\alpha_{\rm s}\alpha_P-1)>0$.
The situation is very similar even in hydrogen shell sources if they are powered by the CNO cycle.
In conclusion, the shell source is generally stable (i.e., $\sigma<0$) under the condition $c^*<0$, i.e. for negative gravothermal specific heat.

\subsection{The case of $c^*>0$}\label{sect_c_pos}

For $c^*>0$, we  have
\begin{displaymath}
\alpha_{\rm s}\alpha_P -1 < 0 ~~.  
%\rm{or} ~~~ \alpha_{\rm s}\alpha_P-1 > \nabla_{\rm ad}\alpha_{\rm s}\alpha_T .
\end{displaymath}
%With the second condition, we have usually $\sigma_{\rm E} > 0$,
%for the same reason discussed in Sect~\ref{sect_c_neg}
%Then, the shell source is 
%unstable when $a_{\rm s}a_P-1 > \nabla_{\rm ad}a_{\rm s}a_T$.
In this case,
we have  $\alpha_T/(\alpha_{\rm s}\alpha_P-1)<0$.
In general, a shell source is prone to thermal instability
when  $\alpha_T/(\alpha_{\rm s}\alpha_P-1)<0$ since any
density decrease due to expansion will lead to a further temperature 
increase (see Eq.~(\ref{eq3})).
However, if the absolute value of $\alpha_T/(\alpha_{\rm s}\alpha_P-1)$
is large enough, a small increase in temperature 
induces such a large decrease in density that 
the nuclear energy generation may be considerably reduced.  
In this case, we have 
$\sigma_{\rm E} < 0$ and the shell nuclear burning is thermally stable.

This can be realized when \at{} becomes very large
or/and when $ |\alpha_{\rm s}\alpha_P-1| \ll 1$.
A large \at{} can be obtained  if radiation pressure
becomes substantial (Fig.~\ref{fig_thermo}c), showing again
its stabilizing effect.
When $ |\alpha_{\rm s}\alpha_P-1| \ll 1$ is satisfied, we have 
$\delta \rho/\rho = \frac{1}{\alpha_{\rm s}} \delta P/P \simeq \alpha_P \delta P/P$.
This means that the shell source behaves as if it were isothermal when 
$\alpha_{\rm s}\alpha_P\simeq1$.
As a consequence, the influence of the temperature perturbation becomes
negligible compared to that of the density perturbation, resulting
in $\sigma_{\rm E} < 0$.
For an ideal gas with $\beta=1$, this condition means $\alpha_{\rm s} \simeq 1$:
the shell source is stable if its thickness $D$ 
is not much less than 0.37$r_{\rm s}$, even when $c^* > 0$ (see Fig.~\ref{fig_crit1}).
This mode of stability has been already noted by Fujimoto~(\cite{Fujimoto82b}):
he found that 
a very large $c^*$ can render a shell source stable,
which is the case when $ |a_{\rm s}a_P-1| \ll 1$ (Eq.~(\ref{eq5})).
As shown in Sect.~\ref{sect:application}, this situation may occur in the
helium shell sources of AGB stars and accreting white dwarfs.

\subsection{Infinitely thin shell sources}\label{sect_infthin}
If a shell source is infinitely thin, we have $\alpha_{\rm s} \rightarrow 0$,
$c^*\rightarrow c_P$ and 
\begin{equation}\label{eq23}
\sigma= \nu - 4 + \kappa_{\rm T} - \alpha_T
(\lambda + \kappa_{\rho}).
\end{equation}
The shell burning can be stable if
the dependence of the nuclear reaction on 
temperature is weak
or/and when radiation pressure is significant,
in which case \at{} becomes  large 
(cf. Fig.~\ref{fig_crit1}).
An astrophysical application for such a case 
is shell burning in accreting neutron stars, for which
the compactness of the gravitating body
enforces a constant pressure in the shell source,
such that $\alpha_{\rm s} \rightarrow 0$.
In fact, the criterion for 
thermal instability in this case 
is very similar to that given by Fujimoto et al.~(\cite{Fujimoto81}), who
studied shell flashes on accreting neutron stars. 
Numerical calculations by Taam et al. (\cite{Taam96}) show that
the helium shell burning in a neutron star can be stable
when the temperature of the helium shell source is as high as  $ 5.6\times 10^8 {\rm K}$,
for which the temperature sensitivity of the $3\alpha$ reaction is 
considerably weakened ($\nu \simeq 5$, Fig.~\ref{fig_nu_3alpha}).  
This can be easily understood by our stability
criterion with Eq.~(\ref{eq23}) 
(see also Fig.~\ref{fig:rs_he} 
and discussions given in Sect.~\ref{sect_ills}).

\subsection{Complete degeneracy}

In completely degenerate gas ($\eta\rightarrow\infty$), we have
$\alpha_T=0$ 
and $\sigma$ becomes 
\begin{equation}
\sigma = \nu-4+\kappa_T.
\end{equation}
Since heat in a strongly degenerate gas is transfered mainly by electron conduction,
with $\kappa_T\simeq2$ (e.g. Kippenhahn \& Weigert~\cite{Kippenhahn90}),
any nuclear reaction with $\nu>2$ will give $\sigma>0$,
resulting in unstable nuclear burning.
An example for the shell burning in strongly degenerate conditions  
is provided by the sub-Chandrasekhar mass progenitor model for
Type Ia supernovae.
In this model, helium ignites at the bottom of a helium
layer of some tenth of solar mass accumulated on a CO white dwarf
with low accretion rates (e.g. Woosley \& Weaver~\cite{Woosley94}). 
At the ignition point, the helium envelope is highly
degenerate ($\eta > 10$, e.g. Yoon \& Langer~\cite{Yoon04b}), 
rendering complete degeneracy as a good approximation. 

\section{Instability diagrams}\label{sect_ills}

\begin{figure}[t]
%\epsfxsize=\hsize
%\epsffile{fig_crit1.eps}
\resizebox{\hsize}{!}{\includegraphics{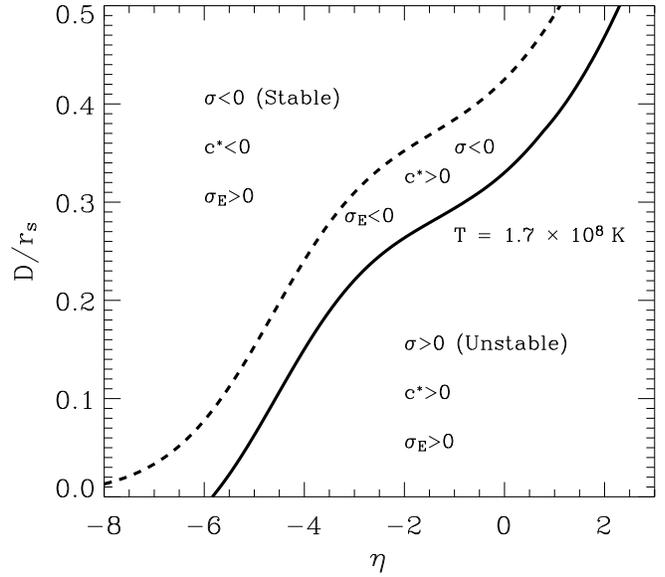}}
\caption{ 
Loci of $\sigma$ and $c^*$ for their sign in $\eta - D/r_{\rm s}$ plane 
when $T=1.7\times10^8 \rm{K}$.
The solid line separates the thermally unstable ($\sigma>0$)
from the stable ($\sigma<0$) region, while the dashed line separates the region with 
negative $c^*$ from that with positive $c^*$.
The same chemical composition as in Fig.~\ref{fig_kapa_deriv} is assumed.
}\label{fig_crit1}
\end{figure}
\begin{figure}[t]
%\epsfxsize=\hsize
%\epsffile{fig_crit2.eps}
\resizebox{\hsize}{!}{\includegraphics{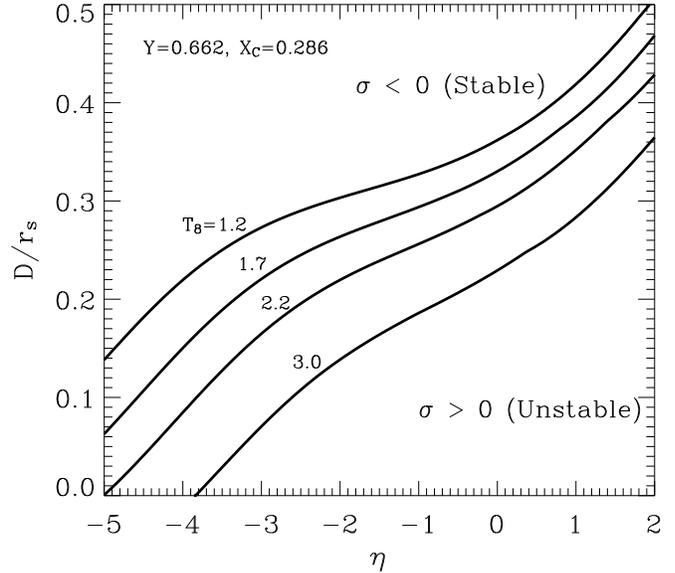}}
\caption{ 
Loci for the sign of $\sigma$  
at 4 different temperatures: $T=1.2, 1.7, 2.2~\rm{and}~3.0 \times 10^8 \rm{K}$ 
in  $\eta - D/r_{\rm s}$ plane.
}\label{fig_crit2}
\end{figure}

\begin{figure}[t]
%\epsfxsize=\hsize
%\epsffile{rs_he.eps}
\resizebox{\hsize}{!}{\includegraphics{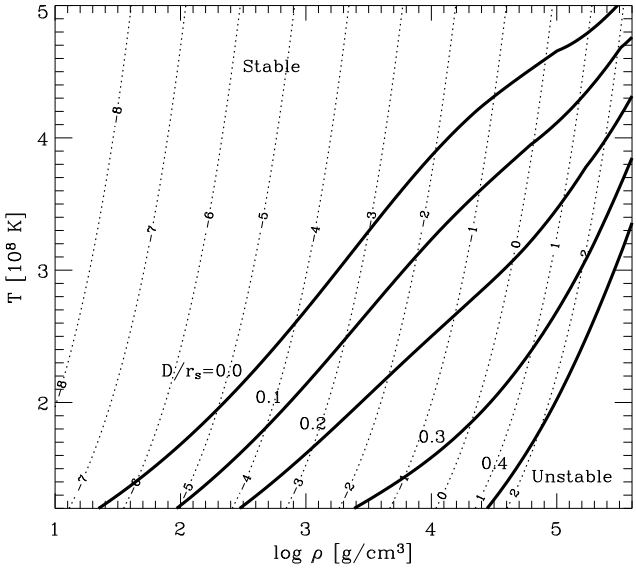}}
\caption{ 
Stability conditions for a helium shell source in the density-temperature plane.
The solid line separates the thermally unstable ($\sigma>0$, lower right part)
from the stable ($\sigma<0$) region, for~5 different relative shell source thicknesses
(i.e., \DR{} = 0.0, 0.1, 0.2, 0.3 and 0.4). 
The dotted contour lines denote the degeneracy  parameter $\eta$. 
$X_{\rm He} = 0.662, X_{\rm C} = 0.280$ and $Z = 0.02$ have been assumed.
}\label{fig:rs_he}
\end{figure}

\begin{figure}[t]
%\epsfxsize=\hsize
%\epsffile{rs_hd.eps}
\resizebox{\hsize}{!}{\includegraphics{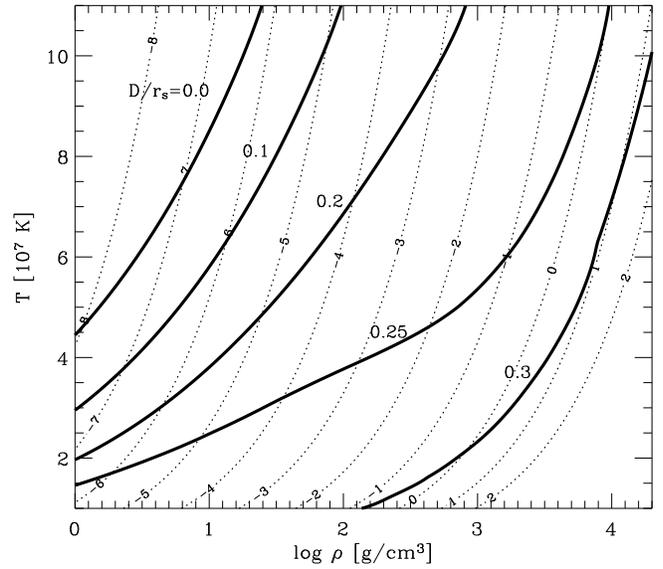}}
\caption{ 
Stability conditions for hydrogen shell source powered by 
the CNO cycle, in density and temperature plane.
The solid line separates the thermally unstable ($\sigma>0$)
from stable ($\sigma<0$) region at 5 different relative thicknesses
of shell source (i.e., \DR{} = 0.0, 0.1, 0.2, 0.25 and 0.3). 
A shell source with each \DR{}  is expected to be 
stable in the upper region of the each locus.
The dotted contour lines denote the degeneracy parameter $\eta$. 
$X_{\rm H} = 0.502, X_{\rm He} = 0.478$ and $Z = 0.02$ have been assumed.
Note that the hot CNO cycle is not considered in the calculation.
}\label{fig:rs_hd}
\end{figure}

Figure~\ref{fig_crit1} gives an example of the stability criterion
for a He shell source in $\eta - D/r_{\rm s}$ parameter space,
using a specific chemical composition as indicated in the figure caption, which
are typical for helium shell sources in AGB stars or in accreting white dwarfs.

This figure shows the two modes 
of stability as discussed in sections~\ref{sect_c_neg} and~\ref{sect_c_pos}:
one with $c^* < 0 ~\&~ \sigma_{\rm E} > 0$ and the other
with $c^* > 0 ~\&~ \sigma_{\rm E} < 0$. 
As expected, a shell source
becomes more susceptible to thermal instability
when it is thinner and more degenerate. 
For $\eta \gtrsim 2$, where
the electron degeneracy is significant ($F(\eta)\gtrsim 1.5$), the shell source 
can not be stable even with $D/r_{\rm s} > 0.5$, 
which is because a pressure change becomes less sensitive to a temperature change
as the electrons become more degenerate. 
On the other hand, the shell source can be stable
even with $D/r_{\rm s} = 0$ when  $\eta \lesssim -5.8$ (see also Fig.~\ref{fig:rs_he}), 
where the radiation pressure is dominant over the gas pressure.

As shown in Fig.~\ref{fig_kapa_deriv}, $\kappa_T$ and $\kappa_{\rho}$
remain to be small compared to $\nu$ in the given temperature and density range.
The quantities \ap{} and \at{} do not
depend strongly on $\mu_{\rm e}/\mu_{\rm I}$.  
The main factor which determines the stability at given $\eta$ and \DR{}
is thus the temperature of the shell source.
Figure~\ref{fig_crit2} shows that the stability 
criterion is more relaxed for higher temperature.
This is because the temperature dependence of the energy
generation is weakened (Fig.~\ref{fig_nu_3alpha}), and because 
radiation pressure becomes more significant at a given
degree of degeneracy as the temperature increases. 
This temperature dependence of the stability
of shell burning explains, why the accretion rates  
which allow steady burning solutions depend
on the mass of accreting white dwarf 
(e.g. Fujimoto~\cite{Fujimoto82b}; Nomoto \& Kondo~\cite{Nomoto91}; Cassisi et al.~\cite{Cassisi98}). 
I.e., higher accretion rates are required to have stable shell burning
for more massive white dwarfs, in which shell sources are thinner. 
This is mainly because shell sources becomes hotter
with higher accretion rates 
if the accreted matter is burned in a steady way. 

In Fig.~\ref{fig:rs_he}, we show the stability conditions
in the density-temperature plane for~5 different relative shell source thicknesses.
As discussed in Sect.~\ref{sect_infthin}, this figure shows that
a shell source can be stable even when it
is infinitely thin (\DR{} = 0), if the temperature is high enough, since
with high temperature the dependence of the energy generation 
on temperature becomes weakened
and the role of radiation pressure becomes important, giving a large \at.
The stability conditions are relaxed as \DR{} increases. 
The conditions for stability in hydrogen shell sources show a
similar behaviour (Fig.~\ref{fig:rs_hd}), where
the shell source is assumed to be powered by the CNO cycle.
I.e., thinner, cooler and more degenerate shell sources are more prone to 
thermal instability. Note, however, that in the calculation, 
we did not consider the so called {\em hot CNO cycle}, which
may be active at high temperatures ($T > 8\times10^7~{\rm K}$). 

\section{Applications}\label{sect:application}

In this section, we apply the stability criterion to various numerical stellar models:
a 3.0~\Msun{} AGB star, helium accreting CO white dwarfs,
a cooling helium white dwarf with a thin hydrogen envelope, 
and a 1.0~\Msun{} giant. 
All models have been computed with a 
stellar evolution code (Langer~\cite{Langer98} and references therein)
which solves
the hydrodynamic form of the stellar structure
equations (Kippenhahn \& Weigert~\cite{Kippenhahn90}).
Opacities are taken from Iglesias \& Rogers (\cite{Iglesias96}).  
We define the thickness of a shell source $D$ such that
the energy generation rate at each boundary
has a certain fraction ($:= f_{\rm b}$) times its peak value.
Values of \DR{} in stellar models would be affected by different
choices of $f_{\rm b}$. For instance, $f_{\rm b} = 10^{-3}$ 
results in a larger \DR{} by about 15 \% compared to
the case of  $f_{\rm b} = 10^{-2}$, in helium accreting white
dwarf models presented in Sect.~\ref{sect_heliumacc}. 
In our applications,
we choose $f_{\rm b} = 2 \times 10^{-3}$ 
because this particular choice gives 
the best prediction for the
onset of thermal instability of shell burning 
for all different types of stellar models 
discussed below. 
The mean values weighted by the energy generation rate 
over the shell source are used
for density, temperature and chemical abundances 
to estimate 
physical quantities which appear in Eq.~(\ref{eq11}).

\subsection{Helium shell burning on the AGB}\label{sect_agb}

\begin{figure}[t]
%\epsfxsize=\hsize
%\epsffile{fig_heshell_1.eps}
\resizebox{\hsize}{!}{\includegraphics{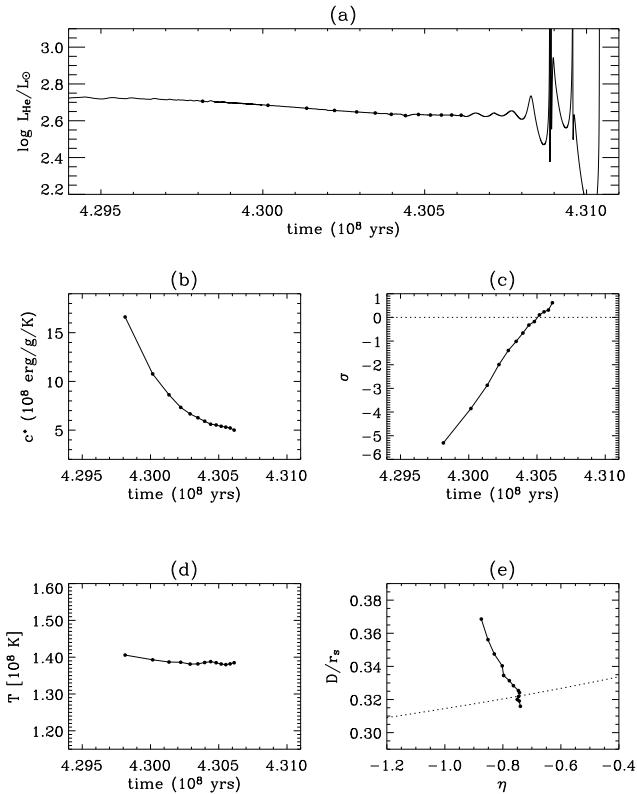}}
\caption{Evolution of the helium shell source in a $3M_{\odot}$ AGB star.
(a) Nuclear luminosity due to helium burning as a function of time, 
for shortly before and during the onset of thermal pulses.
The mean values of $c^*$, $\sigma$, and
temperature in the shell sources of the models marked by a filled circle
are shown in (b), (c) and (d) respectively.
(e) Evolution of the shell source in the $\eta-D/r_{\rm s}$ plane.
The dotted line corresponds to $\sigma=0$ computed for
the physical conditions in the shell source of the last analyzed model.
}\label{fig_heshell_1}
\end{figure}

It is well known that asymptotic giant branch 
(AGB) stars suffer thermally unstable helium shell burning in
their advanced evolution (e.g. Iben \& Renzini~\cite{Iben83a}).  
We investigate this phase by computing a 3~\Msun{} evolutionary model with $Z=0.02$,
starting at the zero age main sequence (Langer et al.~\cite{Langer99}).
After passing through 
hydrogen and helium core burning,
the star enters the helium shell burning stage.
Fig.~\ref{fig_heshell_1}a shows the nuclear luminosity due to helium burning 
as a function of time 
at the onset of the thermal pulses ($T\simeq 4.306 \times 10^8~\rm{yrs}$).
We analyzed those models indicated by a filled circle in Fig.~\ref{fig_heshell_1}a,
in which the helium shell burning is still stable,  
and computed $c^*$ and $\sigma$ for each model.
(Fig.~\ref{fig_heshell_1}b and Fig.~\ref{fig_heshell_1}c). 
During the stable shell burning stage, the mean temperature in the shell source is 
nearly constant at $T\simeq1.4 \times 10^8~\rm{K}$ 
(Fig.~\ref{fig_heshell_1}d).
The shell source becomes thinner and more degenerate 
as the CO core mass increases as shown in Fig.~\ref{fig_heshell_1}e.

The gravothermal specific heat of the shell source 
is positive in the chosen models,
indicating that 
the shell burning is kept to be stable, until the onset of the thermal
pulses by the second mode of the stability discussed in Sect.~\ref{sect_c_pos}.
However, it becomes smaller as the CO core mass increases, 
mainly because \as \ap $- 1$ deviates from 1 more
significantly as \DR{} decreases (see Eq.~(\ref{eq5})), 
and partly because degeneracy in the shell source becomes stronger (Fig.~\ref{fig_heshell_1}e).
The thermal pulses start when the relative thickness of the shell source becomes 
as low as 0.316.

\subsection{Helium shell burning in accreting white dwarfs}\label{sect_heliumacc}

\begin{figure}
%\epsfxsize=\hsize
%\epsffile{fig_heshell_2.eps}
\resizebox{\hsize}{!}{\includegraphics{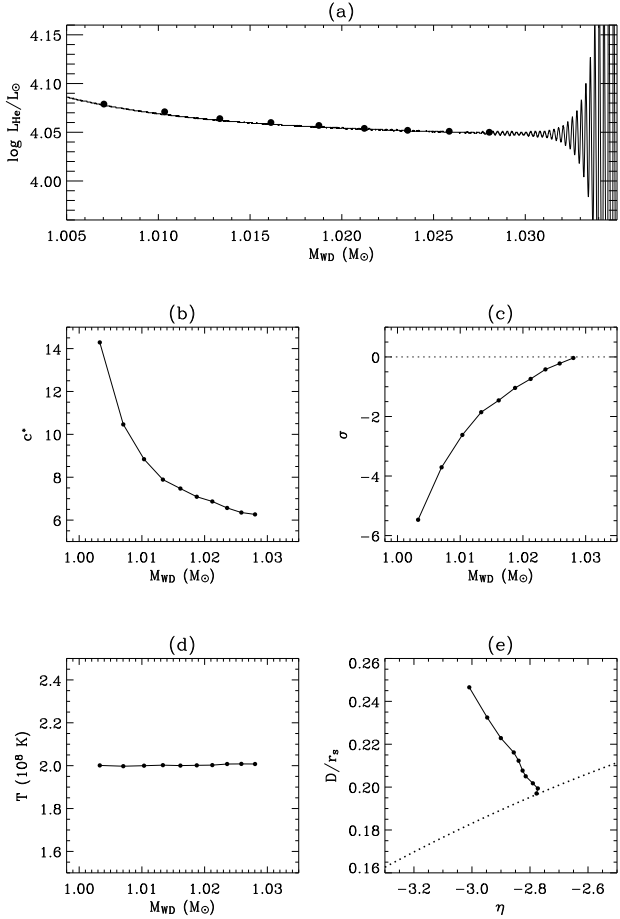}}
\caption{Evolution of the helium shell source in an accreting white dwarf with
an initial mass of $0.998$ \Msun{} and a constant accretion rate of $10^{-6}$ \msyr. 
(a) Nuclear luminosity due to helium burning as a function of the total 
white dwarf mass. 
The mean values of $c^*$, $\sigma$, and
temperature in the shell sources of the models marked by a filled circle  
are shown in (b), (c) and (d) respectively.
(e) Evolution of the shell source in the $\eta-D/r_{\rm s}$ plane. 
The dotted line corresponds to $\sigma=0$ computed for
the physical conditions in the shell source of the last analyzed model. 
}\label{fig_heshell_2}
\end{figure}

Here, we consider the helium shell source in hot helium accreting CO white dwarfs. 
Two different initial white dwarf models are considered: 
\Minit{} = 0.8 \Msun{} with $ \log L_s/{\rm L_{\odot}} = 2.948$
and \Minit{} 0.998 \Msun{} and $\log L_s/{\rm L_{\odot}} = 3.379$, 
where \Minit{} and $L_s$ are the initial mass and surface luminosity, 
respectively.
The accreted matter is assumed to be helium enriched such that $Y=0.98$. 
We have performed simulations with 
6 different combinations of initial mass and accretion rate
(see Table~1).
The helium shell burning is initially stable in each model sequence,
and eventually turns unstable as the white dwarf mass increases 
(see Fig.~\ref{fig_heshell_2} for an example).
The value of $\sigma$ for the shell source 
of the last analyzed model in each sequence
is given in the last
column of the table. 
In all sequences, the last analyzed models,
from which the thermal pulses grow significantly,
have $|\sigma| \lesssim 0.6$,
indicating that our stability criterion predicts the onset of
the instability with good accuracy.

\begin{table}[t]
\begin{center}
\caption{Properties of the computed helium accreting white dwarf models. 
The first column denotes
the model sequence number.
The second column shows the initial mass and
the third gives the accretion rate.
The forth  column gives the amount of accreted
material until the onset of the helium shell instability.
The value of $\sigma$ estimated in the shell source of the last
analyzed model is given in the last column.
}\label{tab1}
\begin{tabular}{c c r c c }
\hline \hline
No.  &   $M_{\rm init}$  (\Msun)  & $\dot{M}$ (\msyr)  &  $\Delta M$  (\Msun) &  $\sigma$ \\
\hline
No.1  & 0.800 & $4.0\cdot10^{-7}$  & 0.0212  & -0.429  \\
No.2  & 0.800 & $5.0\cdot10^{-7}$  & 0.0238  & -0.291  \\
No.3  & 0.800 & $7.0\cdot10^{-7}$  & 0.1141  & -0.598  \\
No.4  & 0.800 & $8.0\cdot10^{-7}$  & 0.1455  & -0.471  \\
No.5  & 0.998 & $8.0\cdot10^{-7}$  & 0.0028  & -0.114  \\
No.6  & 0.998 & $1.0\cdot10^{-6}$  & 0.0048  & -0.035  \\
\hline
\end{tabular}
\end{center}
\end{table}

As an example, results for sequence No.6 are shown in Fig.~\ref{fig_heshell_2}. 
In Fig.~\ref{fig_heshell_2}a, the nuclear luminosity due to helium burning
is given as a function of the total white dwarf mass, which serves as a linear measure of time. 
Thermal pulses are induced 
when the white dwarf mass reaches $\sim 1.028$ \Msun{}. 
The evolution of $c^*$ and $\sigma$ shown in  Fig.~\ref{fig_heshell_2}b 
and~\ref{fig_heshell_2}c
is very similar to that in the AGB model
discussed above. 
The mean temperature in the shell source is about $2\times10^8$ K
(Fig.~\ref{fig_heshell_2}d). 
The shell source becomes
thinner and more dense as the white dwarf mass increases (Fig.~\ref{fig_heshell_2}).
The last analyzed model is at the edge of stability
as shown in Fig.~\ref{fig_heshell_2}c and
the thermal pulses grow significantly from this point.

\subsection{Hydrogen shell burning in cooling helium white dwarfs}\label{sect:hydroshell}

Here, we discuss a hydrogen shell flash 
in a cooling helium white dwarf model (cf., Driebe et al.~\cite{Driebe99}),
which has been obtained by evolving
a 2.4 \Msun{} zero age main sequence star with $Z=0.02$ in a binary
system with the initial orbital period of 1.76 days.
Its companion was a 2~\Msun{} zero age main sequence star.
It loses most of the hydrogen-rich envelope during the hydrogen shell burning stage 
through so-called case B mass transfer, ending in a helium white dwarf
of $M=0.324$ \Msun{} with 
a hydrogen envelope of $M\simeq0.005$ \Msun{} when
the mass transfer stops.
The luminosity and surface temperature at this moment
are $158.5~{\rm L_\odot}$ and $1.4\times10^4$~K, respectively.

\begin{figure}[t]
%\epsfxsize=\hsize
%\epsffile{fig_hewd.eps}
\resizebox{\hsize}{!}{\includegraphics{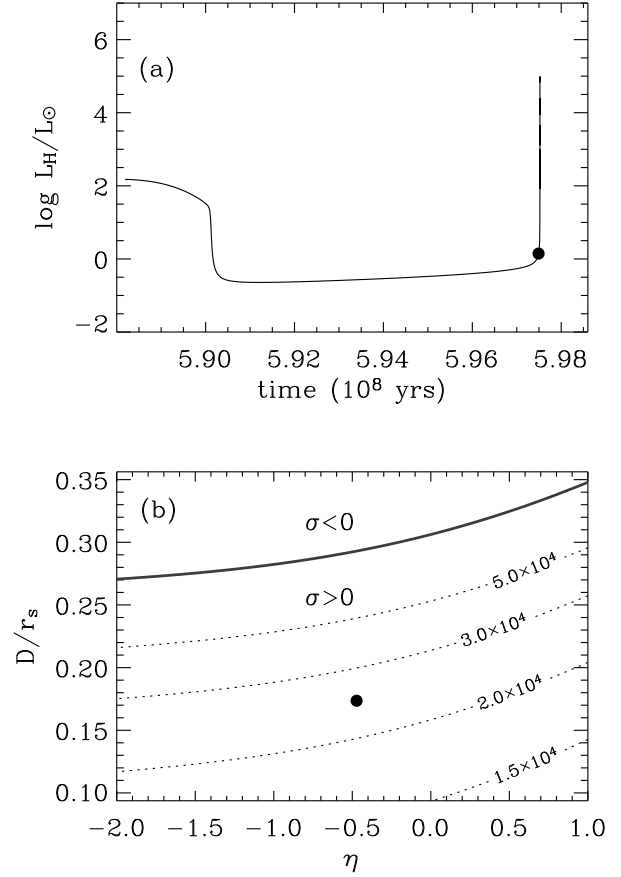}}
\caption{
(a) Nuclear luminosity due to hydrogen burning
at the onset of the hydrogen shell flash in the
hydrogen envelope in a cooling
helium white dwarf of $0.324{\rm M_{\odot}}$. 
(b) Contour levels of the perturbation growth time scale
in the $\eta-D/r_{\rm s}$ plane, 
obtained from Eq.~(\ref{eq12}). Here, $\sigma$ and $\tau_{\rm th}$
have been estimated for the physical conditions
in the shell source of the marked model. The solid
line corresponds to  $\sigma = 0$.
}\label{fig_hewd}
\end{figure}

The hydrogen shell burning
is weakened progressively as the white dwarf cools
(Fig.~\ref{fig_hewd}a).
At $t\simeq5.905\times10^8~{\rm yr}$, 
further weakening of the shell burning stops
due to the increase in density.
The hydrogen burning increases slowly as the helium core
contracts further
and ends in a strong flash at $t\simeq5.98\times10^8~{\rm yr}$.
The thickness and degree of degeneracy of the shell source
have been computed for the model marked by a filled circle
in Fig.~\ref{fig_hewd}a,
and positioned 
in $D/r_{\rm s}-\eta$ plane in Fig.~\ref{fig_hewd}b.
The corresponding temperature and density of the shell source 
are $2.13\times10^{7}$~K and $6.91\times10^2~{\rm g~cm^{-3}}$, respectively.
The shell source at this point is in the unstable regime
as expected from the fact that the nuclear luminosity
reaches its peak very soon thereafter. 
The contour levels for the perturbation growth timescale
from Eq.~(\ref{eq12}) are also given in the same figure.
The perturbation growth time estimated
at the point is about $2.4\times10^4 {\rm yrs}$,
which is in good agreement with the time interval from
the marked model to the peak in $L_{\rm H}$ 
($\simeq 3\times10^4 {\rm yrs}$).

\subsection{Hydrogen shell burning in a 1.0~\Msun{} giant}\label{sect:redgiant}

It is a matter of debate whether
shell flashes in  hydrogen
shell sources of low-mass sub-giants and giants 
could be a possible cause for anomalies of chemical abundances
in stars of globular clusters (e.g. Denissenkov \& van den Berg~\cite{Denissenkov03}). 
Stellar evolution models indicate that hydrogen shell sources in sub-giants
or red giants are usually thermally stable.   
Here, we analyzed the physical conditions of the hydrogen shell source
in a 1.0~\Msun{} model with solar metalicity. 

\begin{figure}[t]
%\epsfxsize=\hsize
%\epsffile{fig_giant.eps}
\resizebox{\hsize}{!}{\includegraphics{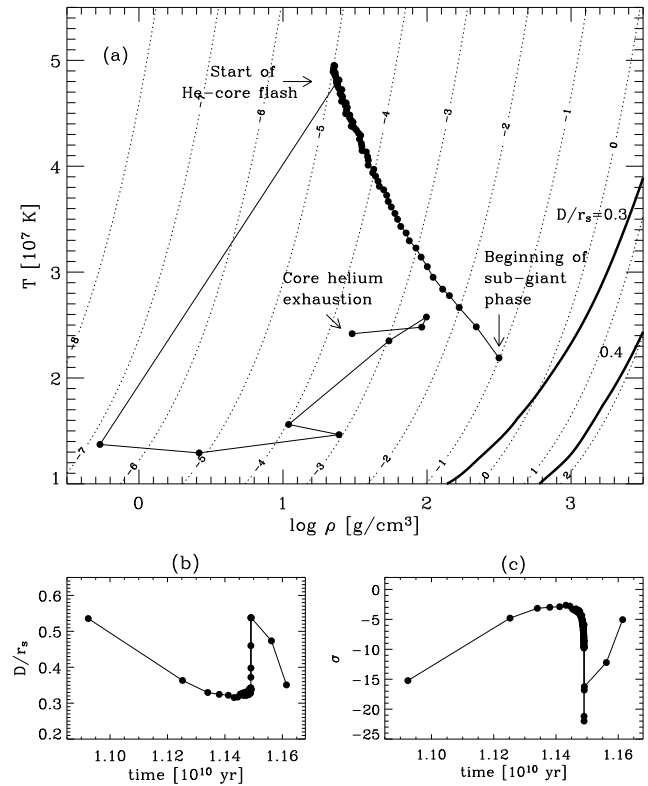}}
\caption{Evolution of a 1.0 star with solar metalicity from the beginning of the sub-giant phase until the
exhaustion of core helium burning.
(a) Evolution of the hydrogen shell source in the density -- temperature plane. 
The thick solid lines correspond to $\sigma=0$ for \DR{} = 0.3 and 0.4 (cf. Fig.~\ref{fig:rs_hd}).
The evolutionary stages at different epochs are indicated with short arrows.
The dotted contour lines give the degeneracy parameter $\eta$.
(b) The relative thickness of the shell source as a function of time. 
(c) The value of $\sigma$ in the shell source as a function of time.
}\label{fig_giant}
\end{figure}

Fig.~\ref{fig_giant}a shows the evolution of the hydrogen shell source 
in the density--temperature plane,
from the beginning of the sub-giant phase until
the core helium exhaustion.
Note that, from the beginning of the sub-giant until
the core helium flash,
the shell source becomes hotter and less dense (Fig.~\ref{fig_giant}a),
while the relative thickness of the shell source remains larger than 0.3 (Fig.~\ref{fig_giant}b).
Once the core helium flash occurs, both the temperature
and density of the shell source decrease rapidly due to the expansion.
Throughout the considered stages, 
the evolutionary track of the shell source remains in the thermally stable regime
(cf. Fig.~\ref{fig_giant}c).
This result indicates that
the onset of the thermal instability in hydrogen shell sources
of low-mass sub-giants and giants is unlikely.

\section{Conclusion}

We have developed a stability criterion for thermonuclear shell sources
which can be easily applied to numerical stellar models in
a quantitative way.
The stabilizing/destabilizing factors have been identified 
unambiguously: a shell source is less prone to
the thermal instability when it is thicker, less degenerate and hotter 
(Sect.~\ref{sect_physcond}, Figs.~\ref{fig_crit2}, \ref{fig:rs_he} and \ref{fig:rs_hd}).  
If a shell source is sufficiently
thick at a given degree of degeneracy, 
the gravothermal specific heat becomes negative and 
any excessive energy gain is consumed mostly for the expansion work,
preventing thermal runaway. 
It is shown that the shell source can remain stable even
with a positive gravothermal specific heat
as long as temperature at a given 
degree of degeneracy is high enough such that
the sensitivity of nuclear reactions to temperature is significantly
weakened or/and that radiation pressure becomes substantial. 
In such a case, the energy
loss rate by radiation begins to dominate over the
additional nuclear energy production rate.
Interestingly, this effect allows even an infinitely 
thin shell source to be thermally stable 
(Sect.~\ref{sect_infthin}, Fig.~\ref{fig:rs_he}
and~\ref{fig:rs_hd}), which may have interesting
consequences in accreting neutron stars 
(Sect.~\ref{sect_infthin}; cf. in't Zand et al.~\cite{Zand03}).

We have applied the stability criterion developed in this study
to a 3~\Msun{} AGB star, to helium accreting CO white dwarfs, to a helium
white dwarf with a thin hydrogen shell, and
to a 1.0~\Msun{} giant  (Sect.~\ref{sect:application}). 
In spite of the simple approach of using homology assumption in 
the shell source and not relying on the full linear stability analysis,
it is demonstrated that this criterion can predict the onset of 
thermal pulses or flashes with a reasonably good accuracy.

In conclusion, being simple and robust, 
our stability criterion may serve as a useful
tool for analyzing shell burning 
in various types of stars.
One example can be found in Yoon et al. (\cite{Yoon04c}),
who discuss the effects of rotationally induced chemical mixing
on the behaviour of helium shell burning in accreting white dwarfs.

\begin{acknowledgements}
We are grateful to the anonymous referee 
for fruitful and enlightening comments which led to improvement
and expansion of the text.
This research has been supported in part
by the Netherlands Organization for
Scientific Research (NWO).
\end{acknowledgements}

\end{document}